\documentclass{article}

\usepackage{tikz} 
\tikzset{mynode/.style={font=\footnotesize,inner sep=0pt,text=black}
}
\usepackage{amsmath}

\usepackage{amssymb}

\begin{document}




\title{A framework for trustworthiness assessment\\based on fidelity in cyber and physical domains}


\author{Vincenzo De~Florio${}^\dagger$ and Giuseppe Primiero${}^\ddagger$\\
${}^\dagger$: MOSAIC, University of Antwerp \& iMinds, 2020 Antwerpen, Belgium\\
${}^\ddagger$: Department of Computer Science, Middlesex University, London, UK}






\maketitle
\begin{abstract}
We introduce a method for the assessment of trust for $n$-open systems based on a measurement of fidelity and present a prototypic implementation of a complaint architecture. We construct a MAPE loop which monitors the compliance between corresponding figures of interest in cyber- and physical domains; derive measures of the system's trustworthiness; and use them to plan and execute actions aiming at guaranteeing system safety and resilience. We conclude with a view on our future work.
\end{abstract}



\section{Introduction}\label{intro}

Fidelity of an open system can be interpreted as the compliance between corresponding figures of interest in two separate but communicating domains, see~\cite{DF14a}. 
In cyber-physical systems, perfect fidelity means that actions in the physical domain 
have a well-defined and stable counterpart in the cyber domain, and vice-versa. This is an ideal state,
as no concrete cyber-physical system may guarantee at all times a perfect correspondence between its domains of action. In practice, fidelity is affected by circumstances that let the system drift from the optimal case.
%
%
%
%
Our stance is that, by observing the characteristics of said drifting, we may introduce a fine-grained characterisation of the quality of system trustworthiness. 
To this aim, we introduce a practical method for the assessment of trust based on the measurement of fidelity in computational systems, including cyber-physical ones. As a way to measure fidelity drifting we propose to adopt and extend the approach described in~\cite{DFB12b,DB07a}.
We propose to make use of the accrued information to assess the characteristics of drifting in fidelity; we derive from it dynamic properties of both user and machine; evaluate it in terms of system's trustworthiness and use it to execute safety-assurance actions. This generates a trust-induced MAPE-loop.


The paper is structured as follows. In Sect.~\ref{s:fidel} we describe the conceptual model of system fidelity; in Sect.~\ref{s:janus} the model for fidelity evaluation is implemented in an architecture for a cyber-physical system;
in Sect.~\ref{s:trust} we build the link to trustworthiness evaluation; finally, in Sect.~\ref{s:end}
we conclude by drawing some observation and describing further lines of research.

\section{Conceptual understanding of System Fidelity}\label{s:fidel}

In the present section we introduce a conceptual model of system fidelity, derived from the one presented in~\cite{DF14a}. Our starting point is the concept of $n$-open systems ($n$OPS), characterised by the following properties:

\begin{itemize}
\item $n$OPS interact with one or more of the environments they are deployed in.
\item $n$OPS base their action on the ability to sense $n$ classes of raw facts taking place
      in their deployment environments.
\item $n$OPS are able to construct and maintain $n$ classes of 
      internal representations of the raw facts, called \textit{qualia}.
\item \textit{Qualia} are, to some extent, faithful, meaning that they timely
      reflect the dynamic variation of the corresponding class of raw facts.
\end{itemize}

We discuss here \emph{fidelity\/} as a characterisation of the above-mentioned faithfulness.
More formally, given any $S\in n$OPS, we consider $n$ classes of raw facts, $[r]_i, 1\le i\le n$, and $n$
classes of binary operations, $[\dotplus]_i$, such that each of the couples 
\begin{equation}
([r]_i, [\dotplus]_i)\label{eq:algstruct}
\end{equation}
constitutes an algebraic structure; for instance, when $[\dotplus]_i$ is a singleton,
then $([r]_i, [\dotplus]_i)$ is a group.
Likewise, for any $1\le i\le n$, we call $[q]_i$ the class of \textit{qualia\/} corresponding to $[r]_i$ and
$[\oplus]_i$ the class of binary operations corresponding to $[\dotplus]_i$.
Moreover, as we did for~\eqref{eq:algstruct}, we assume that $([q]_i, [\oplus]_i)$ is an
algebraic structure. Then, for each $1\le i\le n$, we consider the following function:

\begin{equation}
\Phi_i : [r]_i \to [q]_i,\label{eq:Phi}
\end{equation}
mapping the qualia corresponding to any raw fact in $[r]_i$. We refer to the $\Phi_i$ functions as the \emph{reflective maps\/} of some $n$-open system $S$. Reflective maps are assumed to be bijective functions, with $\Phi^{-1}_i$ being the \emph{inverted reflective maps\/} of $S$ associating the raw fact corresponding to each input quale. We shall say that $\Phi_i$ expresses \emph{perfect fidelity\/} between $([r]_i, [\dotplus]_i)$ and $([q]_i, [\oplus]_i)$ if and only if $\Phi_i$
preserves its algebraic structures (i.e., it is an isomorphism). More formally, for any couple of raw facts $(r_1, r_2)\in [r]_i \times [r]_i$
and for all $\mathbb{+}\in[\dotplus]_i$ and all $\mathbb{\cdot}\in[\oplus]_i$:
\(
        \Phi_i(r_1    +   r_2) = \Phi_i(r_1)  \cdot  \Phi_i(r_2).\label{eq:HiFi}
\)

Perfect fidelity may be better understood through an example. Let us assume that $S$ is a cyber-physical system responsible for the operation of a mission critical hard-real-time service. An operator is responsible for the issuing of requests for service, which is done through a user interface (UI). A set of raw facts and prescribed behaviours pertaining to the physical environment are represented as ``cyber-qualia'' and ``cyber-behaviours'' stored in computer memories. Likewise, a set of ``UI-qualia'' and ``UI-behaviours'' are respectively rendered and operable through the UI. Perfect fidelity states that the correspondence between the physical, the cyber, and the UI domains is such that the prescribed behaviours as well as the referred raw facts and qualia are consistent on either of the involved domains. Thus certain operations and objects represented and rendered via the UI perfectly correspond to operations and objects encoded in $S$'s computer components, which in turn perfectly correspond to physical actions having effects on physical entities. Obviously, perfect fidelity only represents a reference point and can not be sustained and guaranteed at all times in real life. A slightly different and more practical definition of fidelity is given by a function $\phi_i, 1\le i\le n$:

\begin{equation}
\phi_i : [r]_i \to [q]_i, \ \ \hbox{such that} \ \ \ \forall \,+\in[\dotplus]_i, \forall \,\cdot\in[\oplus]_i:
        \phi_i(r_1    +  r_2) = \phi_i(r_1)  \cdot  \phi_i(r_2)  \cdot   \Delta_i(t).\label{eq:phi}
\end{equation}
As for the $\Phi_{i}$ function, $\phi_{i}$ returns the qualia associated with the input raw facts. Contrarily to the $\Phi_{i}$ function, the $\phi_i$ does not preserve
their algebraic structures unless the value of the error component $\Delta_i(t)$ is zero. The use of lower-case
``$\phi$'' is meant to suggest that $\phi_{i}$ represents a less-than-perfect version of $\Phi_{i}$.  The $\Delta_i(t)$ quantifies a drifting in time (represented
by variable $t$) of the ability to create a trustworthy ``internal'' representation of an experienced raw fact.

In~\cite{DF14a}, fidelity is classified in function of the type of drifting. Classes may include, e.g., the following cases:

\begin{itemize}
\item Hard-bound fidelity drifting, exemplified by hard-real-time $n$OPS.
\item Statistically-bound fidelity drifting---as typical of, e.g., soft real-time systems.
\item Unbound fidelity drifting characterised by a ``trend''.
\item Unbound fidelity drifting with a random trend.
\end{itemize}

Accordingly, very disparate cases can be presented to exemplify imperfect fidelity, e.g. the accidents experienced by the linear accelerator Therac-25~\cite{Therac93,De10} and the system failure caused by the last Scud fired during the Gulf War~\cite{PatriotNYT}. In the former case, one of the reasons that led to some of the accidents was that the UI-qualia and the cyber-qualia did not match when the operator was typing at a very fast pace. As a result of such imperfect fidelity, the quantities represented on the screen of the Therac-25 did not correspond to the data stored in its memories---and, regrettably, to the amount of radiations supplied to the patients by the linear accelerator. In the Patriot case, resulting in 28 US Army reservists being killed and 97 injured by a Scud missile on 25 February 1991. The missile-defence system was an $n$OPS that interacted with its environment through a number of context figures that included velocity and time. As discussed in~\cite{Assets:GrTr07}, the cyber-quale corresponding to physical time was represented as the number of tenths of seconds from a reference epoch and stored in a 24-bit integer variable. Imprecision in the conversion of said variable into a real number translated in an \emph{unbound drifting of fidelity over time}. The more the Patriot missile-defence system operated without a reboot, the larger was the $\Delta_{i}$ pertaining to time and, as a consequence, the greater the discrepancy between the expected and the real position and velocity of the incoming Scud missile.  An obvious workaround to the above unbound drifting is that of rebooting the system regularly so as to rejuvenate~\cite{GRT} the qualia management system and bring back the $\Delta_{i}$ to ``safe'' values. Although both problem and workaround were known at the time of the accident, no upper bound was known beyond which the resilience of the system would be affected. Common belief was that the unresilience threshold would never be reached in practice. Regrettably, reality proved the trust on that belief to be misplaced.
The Patriot missile that had to intercept the Scud never took off. The cases of the Therac-25 and of the Patriot system reveal a common denominator: behaviours such as those of a human operator or those produced by a numerical algorithm are all translated into a same, homogeneous form: that of a stream of numerical data
representing samples of the $\Delta_i(t)$ dynamic systems. A major methodological assumption in the present work is that the above data could be compared with other data representing reference conditions. In the Therac-25 case, such data may correspond to, e.g., reference user stereotypes of expected operator behaviours, represented for instance as the numerical weights in a Hidden Markov Model~\cite{VD13}. Likewise, in the Patriot case, those reference data may correspond to, e.g.,
a threshold representing safe ranges for  accumulated or cumulative numerical errors produced through the iterations of numerical methods. In both the exemplified cases, assessing fidelity is thus translated into the problem of evaluating a ``distance'' between observed and reference data.

In what follows, we propose an architecture and a prototypical system to evaluate systematically a system's fidelity drifting. These are modelled on a toy example, easily modified and extended to a real case scenario, whose presentation we reserve to an extended version of this work.

\section{Janus' Architecture}\label{s:janus}

\begin{figure}
\includegraphics[width=1\textwidth]{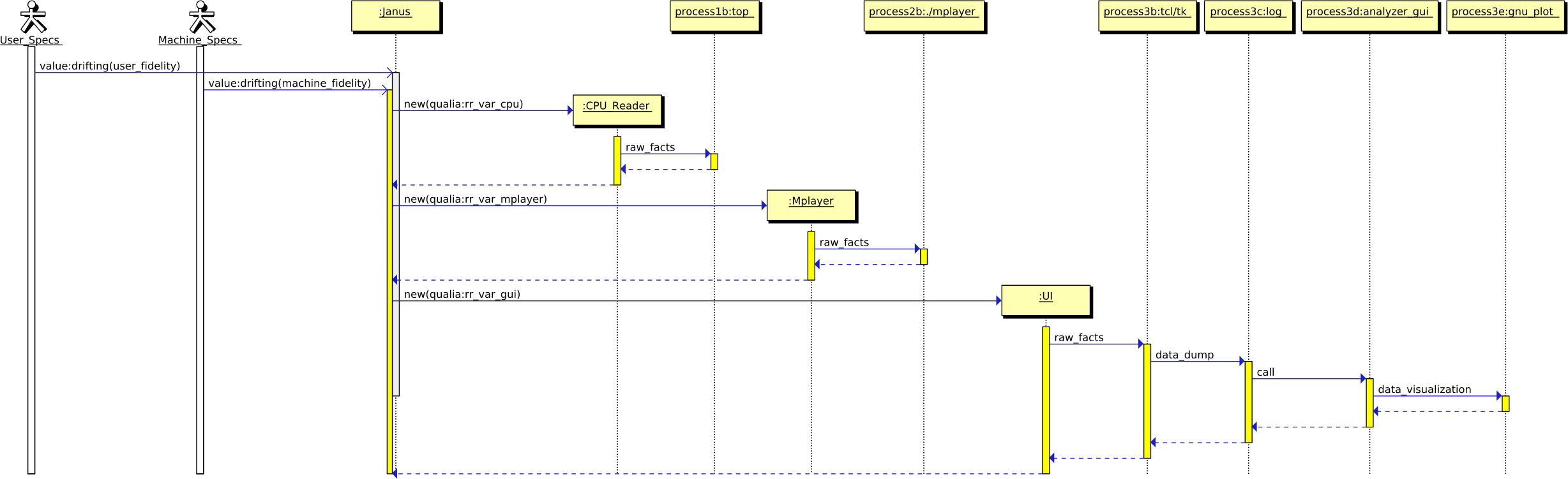}
\caption{Sequence diagram of the Janus' components.}
\label{f:Janus}
\end{figure}

In view of the above discussion, our approach to systematic evaluation of fidelity requires at least the following
components:

\begin{itemize}
\item A sensory service, interfacing the deployment environments so as to register
a number of ``raw facts'', namely
variations in a number of environmental properties. Raw facts could refer, for instance, to variations
in luminosity, temperature, or the amount of network bandwidth available between two
endpoints.
\item A uniform qualia service, providing consistent, timely and reliable access to 
cyber-qualia (computer-based representations of the raw facts).
\item An application layer, providing a convenient means for fidelity assessment and reactive control.
\end{itemize}
%
In what follows we introduce the above components and a prototypical system
compliant to the just sketched models, see Fig.~\ref{f:Janus}.

\subsection{Sensory and Qualia Services}\label{s:janus:rrvars}
Our sensory and qualia service is based on so-called reflective and refractive
variables (RRvars)~\cite{DeBl08a,DB07d}, a tool for the development of $n$OPS in the C
programming language. The idea behind RRvars is quite simple: a number of
predefined variables provide access to the qualia of corresponding raw facts.
Those variables are ``volatile'', meaning that their content is asynchronously
and continuously updated by an associated thread. Thus for instance RRvar \verb"int cpu" does not
retain a constant value; rather, it is continuously updated by the \verb"Cpu()" thread.
Such thread cyclically
retrieves the percentage of CPU currently in use and stores it in the
memory cells associated to \verb"cpu" as an integer number ranging between 0 (CPU fully available)
and 100 (no CPU available). In the current implementation, which runs on 
Linux and Windows/Cygwin systems, \verb"Cpu()" retrieves its raw facts by calling
the \verb"top" utility.
This is referred to as \texttt{process1b} in Fig.~\ref{f:Janus}.

A second and slightly more complex example is given by RRvar \verb"int mplayer", a variable updated
by thread \verb"Mplayer()", referred to as \texttt{process2b} in Fig.~\ref{f:Janus}. The latter component communicates with an instrumented MPlayer movie player~\cite{Mplayer} through
a simple UDP client/server protocol. By reading the content of \verb"mplayer" one is informed of the state of the MPlayer---see Fig.~\ref{f:newJanus1}.
Currently, the following integer values are used:
\begin{verbatim}
#define UDPMSG_STOP     1 // mplayer has finished playing a video
#define UDPMSG_SLOW     2 // mplayer is encountering problems while playing a video
#define UDPMSG_PAUSED   3 // mplayer has been paused
#define UDPMSG_START    4 // mplayer has been launched
#define UDPMSG_SIGNAL   5 // mplayer caught an exception and is about to exit abnormally.
\end{verbatim}

A third case is given by RRvar \verb"int ui", updated
by thread \verb"Ui()". This is a special case in that this RRvar represents a UI-qualia (see Sect.~\ref{s:fidel})
reporting raw facts specific of an instance of a user interface. This is referred to as \texttt{process3b}--\texttt{process3e} in Fig.~\ref{f:Janus}.
Said user interface and the 
\verb"Ui()" thread communicate transparently of the operator via the same mechanism presented for
RRvar \verb"mplayer". The values returned in RRvar \verb"int ui" represent usability raw facts
derived by comparing the behaviours exercised by the current user with ``reference behaviours''
representing the expected behavioural patterns of a trustworthy operator. The method to derive
these raw facts is described in~\cite{DFB12b,VD13}. This method may be used
to detect gradual behavioural driftings (due to, e.g., fatigue, stress, or the assumption of psychotropic substances) and sudden behavioural driftings (caused, e.g., by an account takeover or other cyber-criminal attacks).


\subsection{Control Layer: Janus}\label{s:janus:janus}

\begin{figure}
\centerline{\includegraphics[width=0.5\textwidth]{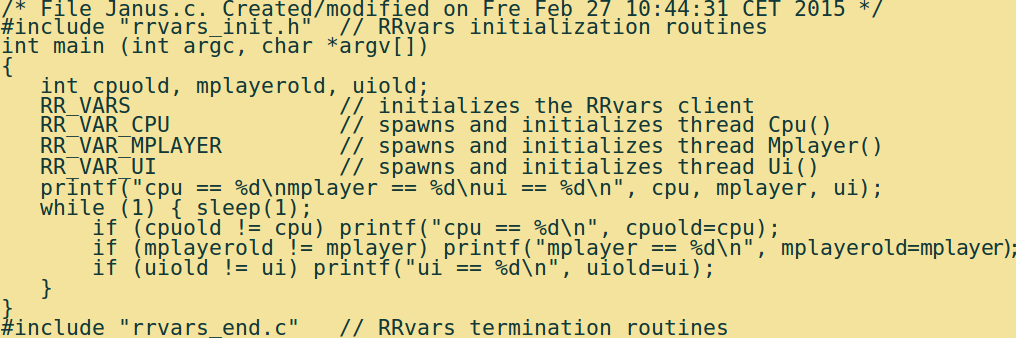}}
\caption{Typical structure of an RRvar client. Here three RRvars (cpu, mplayer, and ui) are declared and continuously
displayed.}
\label{f:Janus1}
\end{figure}

\begin{figure}
\centerline{\includegraphics[width=1.0\textwidth]{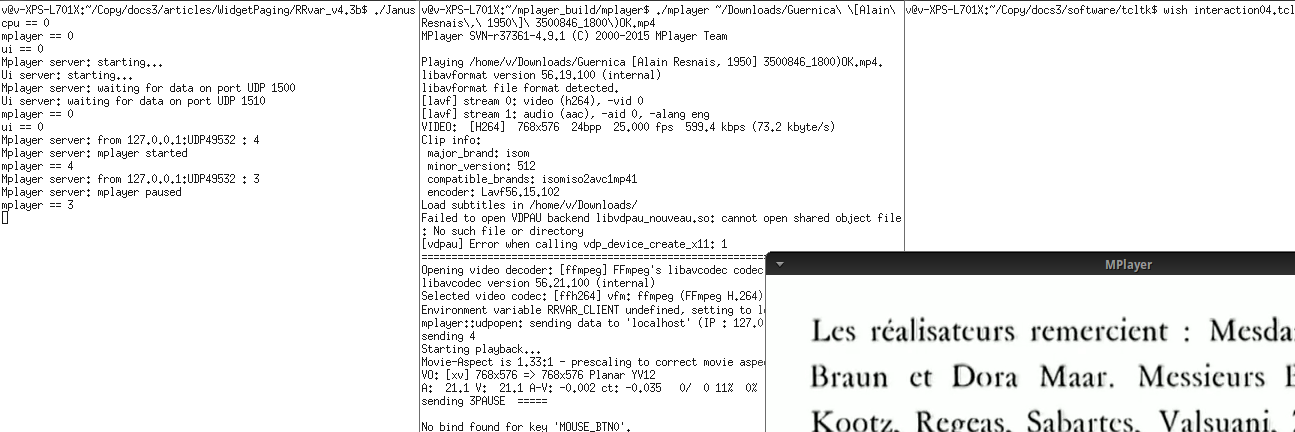}}
\caption{An instance of the MPlayer connects with Janus and reports its state.}
\label{f:newJanus1}
\end{figure}

\begin{figure}
\centerline{\includegraphics[width=1.0\textwidth]{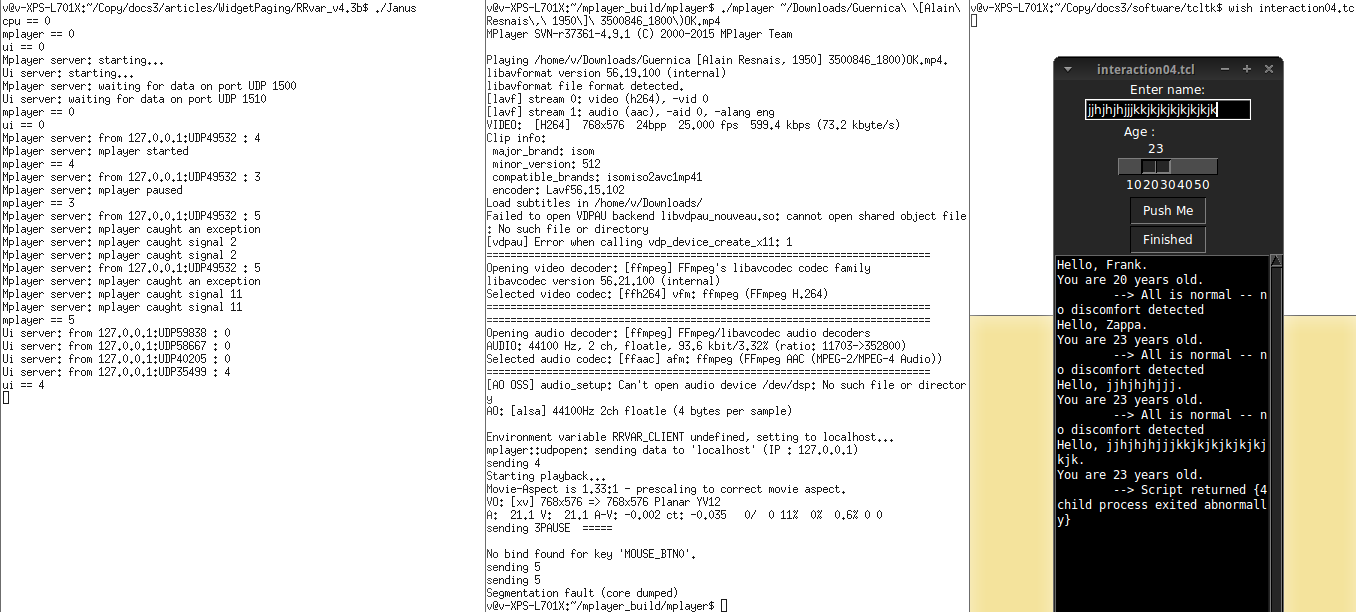}}
\caption{The RRvar client connects with both MPlayer and an exemplary user interface.}
\label{f:newJanus2}
\end{figure}

Janus is the name of our exemplary  RRvar client component.\footnote{As for the
system described in~\cite{DeDe98}, the name of our component comes from
mythical \emph{Janus Bifrons}, the god of transitions, who had two faces and thus could observe
and reason by considering two different ``views'' at the same time. Of the proposed etymologies
of Janus, particularly intriguing here is the one proposed by Paul the Deacon~\cite{Festus}:
hiantem, hiare, ``to be open''. Due to this fact one would be tempted to refer to Janus Bifrons here as to a \emph{2-open system}.} The structure of \verb"Janus" is the one typical of RRvar clients~\cite{DB07d} and exemplified in Fig.~\ref{f:Janus1}.
As can be seen from the picture, the RRvar metaphor makes it possible to quickly define $n$OPS components
based on the three classes of qualia presented above, see Fig.~\ref{f:newJanus2}.

\begin{figure}
\centerline{\includegraphics[width=1.0\textwidth]{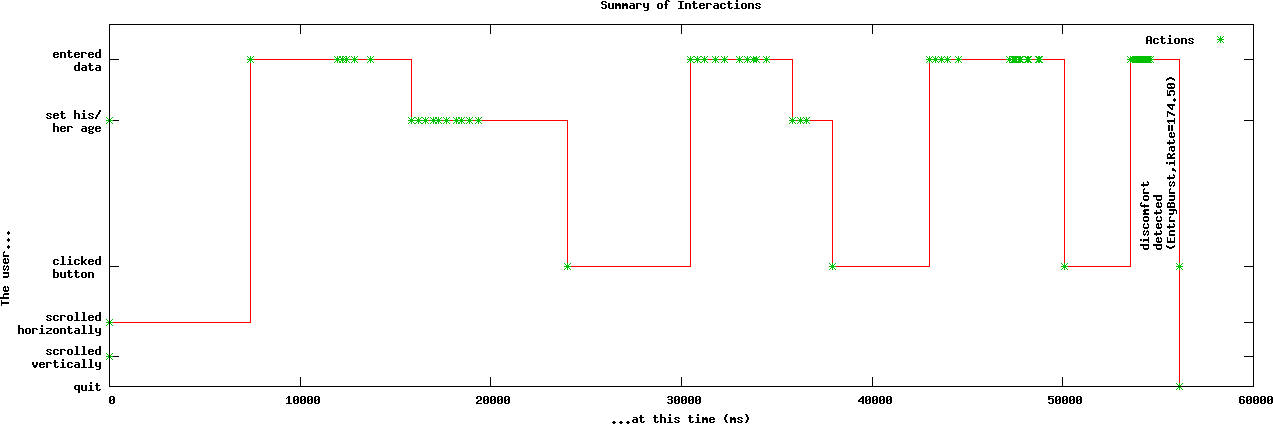}}
\caption{Log of the interaction between user and UI. Between 50 and 60s a rapid burst of keystrokes
is interpreted as an anomalous situation.}
\label{f:interaction}
\end{figure}

\section{Fidelity as Trustworthiness}\label{s:trust}

\begin{figure}
\centerline{\includegraphics[scale=0.45]{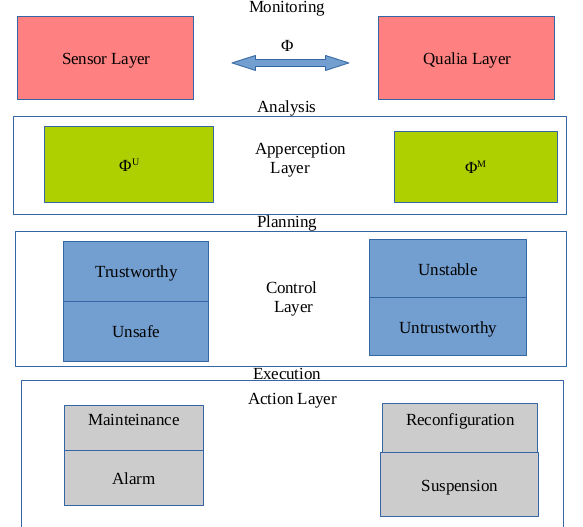}}
\caption{A MAPE-loop for System trustworthiness based on fidelity.}
\label{fig6}
\end{figure}

In this final section, we link the model of open systems offered in Sect.~\ref{s:fidel}, and their architecture implemented in Sect.~\ref{s:janus}, to a model of trustworthiness evaluation. We use fidelity to assess the working conditions of an open system and to establish a metric of trustworthiness. Our final objective is to provide a qualitative extension of the standard MAPE-loop based on trustworthiness assessment. Trust for security, management and reputation systems is gaining a lot of attention in the literature and it is typically accounted for as a first-order relation between (possibly autonomous) agents or system's modules, see e.g. \cite{eps262294,ClarkeCX09,as-trust2011,5340804,journals/tdsc/YanP11}. Our account considers trust as a second order property characterising cumulatively system's fidelity, where the latter is obtained as the dynamic variation of properties of the system's modules. This approach to second-order trust has been already used for information transmission evaluations (\cite{primiero_taddeo}), access-based control (\cite{primiero_secureND}), and software management systems (\cite{primiero_NDC}). In the present analysis, trustworthiness is used to plan reaction to malfunctioning and restoring of functionalities in cyber-physical systems. As the cases of Therac-25 and the Patriot-missile defence system show, unacknowledged drifting can be crucial in maximising unjustified trust to dangerous levels. On the other hand, a metric evaluating minimal functionality thresholds can minimise unjustified mistrust, reducing confidence that the system \textit{will} choke and eventually fail (antitrust). Early cases of low true-alarm rate or high false-alarm rate in automation do not need to set constants for future behaviour of the user. Similarly, criteria to compare intended and current behaviour are essential to allow mechanical assessment of user's inability, incompetence or threats. Fidelity and drifting can provide the systematic methodology and quantitative model to continually evaluate and experiment the system's vulnerability in the context of its operations, thus making the dichotomy conditional vs. unconditional trust (faith) working. Trust can be thought of as a measure of confidence about minimal drifting from fidelity for all the components, and hence that the overall behaviour of the system is sufficiently resilient. In the following, we reconstruct the process of fidelity monitoring, trustworthiness evaluation and operation execution in terms of a MAPE-loop designed for trust, see \cite{MAPE}. In the architecture presented in Sect.~\ref{s:janus}, the \textit{Janus} assesses the behaviour of the system. This is parametrised in view of reflective variables for CPU consumption and a component's operations (on the machine side of the system), and for the user interface (on the user's side). Mapping of these variables values as raw facts and qualia provides a measure of system's fidelity:

\begin{itemize}
\item $\Phi_{CPU}: [r]_{CPU} \to [q]_{CPU}$, obtained by the mapping of input values from the $top$ process to pre-selected parameters assigned to CPU-consumption behaviour;
\item $\Phi_{component}: [r]_{component} \to [q]_{component}$, obtained by the mapping of input values from the executable's observable behaviour to pre-selected parameters assigned to its operations;
\item $\Phi_{UI}: [r]_{UI} \to [q]_{UI}$, obtained by the mapping of input values from  the user's observable behaviour to pre-selected parameters assigned to a standard or expected user's behaviour.
\end{itemize}
The system's component monitoring the classes $[r]_{i}$ of raw facts is called the \textit{sensor layer}; similarly, we use \textit{qualia layer} to refer to the component monitoring the classes $[q]_{i}$ of qualia. The combination of the sensory and representative layers constitutes the \textit{Monitoring} component within our MAPE-loop. Fidelity is then approximated as the inversely proportional function of the drifting from appropriate mappings $\Phi_{i}$. We shall refer to the value of user-based mappings as \textit{user-defined fidelity} ($\Phi^{U}$); correspondingly, we shall call \textit{machine-defined fidelity} ($\Phi^{M}$) the value based on mappings related to the machine behaviour. For the \textit{Janus} introduced in Sect.~\ref{s:janus}, 

\begin{equation}
\Phi^{U}= 1/\Delta(t)_{UI}
\end{equation}
\begin{equation}
\Phi^{M}= 1/f(\Delta(t)_{CPU},\Delta(t)_{exec})
\end{equation}
where $f$ is some function, weighted according to domain-specific and user defined parameters. We refer to the set of values $\mathbf{\Phi}(t)=\{\Phi^{U},\Phi^{M}\}$ as the content of our \textit{Analysis} component, with the global value $\mathbf{\Phi}$ parametrised by time. As an example, consider the class of mappings $\Phi_{UI}$, with a value of the sensor layer indicating e.g. quick typing and a value of the qualia layer returning a distress indication: in this case the fidelity layer reports an high value. As an example across distinct mappings, assume that the reflective variable for MPlayer indicates that the application is running slower, while the one for CPU monitoring indicates low usage value: this is expressed by a low fidelity value across the two classes in $\Phi^{M}$. Analysing fidelity values across the distinct monitoring layers allows a cumulative evaluation to be obtained. This value is monitored by the \textit{apperception layer}. This layer is used to evaluate system trustworthiness as a global value of user-defined and machine-defined fidelity values.
The next level is represented by the \textit{Janus} feeding the content of the \textit{apperception layer} into the \textit{control layer}. This corresponds to the \textit{Planning} component of our system.
%
%
%
%
%
At this stage, system trustworthiness is matched to a resilience scale that identifies and automatically triggers actions aimed at preserving system safety or enabling ameliorating conditions. The latter part of the system is the \textit{Execution} component, monitoring an \textit{action layer}.
%
Despite the fact that a resilience scale should be highly domain specific, a possible general model can be given in terms of four essential stages:

\begin{enumerate}
\item \textit{Trustworthy System} identifies high levels of $\mathbf{\Phi}(t)$, inducing optimal, sustainable working conditions;
\item \textit{Unstable System} identifies high-to-medium $\Phi^U$ and low $\Phi^M$ levels, inducing reconfigurable working conditions;
\item \textit{Unsafe System} identifies high-to-medium $\Phi^M$ and low $\Phi^U$ levels, inducing alarm-rising working conditions;
\item \textit{Untrustworthy System} identifies low-levels of $\mathbf{\Phi}(t)$, inducing inadvisable or below-safety working conditions.
\end{enumerate}
The above analysis is summarised in the MAPE-loop in Fig.~\ref{fig6}.

%
%

\section{Conclusions}\label{s:end}

We have presented a prototypical model architecture for a trust-based MAPE-loop for cyber-physical systems. The trust evaluation is grounded on the assessment of fidelity drifting with respect to values representing
ideal reference conditions for both user and machine.
Although prototypical, we believe that our architecture proves the feasibility of our approach to trustworthiness assessment
and paves the way towards future implementations.
Our goal is to develop systems that are trustworthy in integrating quality-of-service and quality-of-experience, by optimising the  relations between system-level, ``microscopic'' aspects and user-level, ``macroscopic'' ones.
A further goal is to extend our architecture into that of a MAPE-K loop and apply machine learning methods
such that systems based on our approach may systematically improve the match with the environments
they interact with.
Costs analysis for the trust-based MAPE-loop remains to be explored. Its use at early stages of design can reduce risks; our examples shows, nonetheless, that it is possible to deploy the methodology on existing software, by selecting relevant variables. Further work will focus on formal modelling of  trustworthiness assessment in a probabilistic epistemic setting.

%



%


\bibliography{thesis}

\begin{thebibliography}{10}

\bibitem{IMTEC92:26}
Anonymous.
\newblock Patriot missile defense: Software problem led to system failure at
  dhahran, saudi arabia.
\newblock Technical Report GAO/IMTEC-92-26, U.S. Government Accountability
  Office, US, General Accounting Office, Washington, D.C., 20548, 1992.

\bibitem{Mplayer}
Anonymous.
\newblock Mplayer --- the movie player, 2015.
\newblock Retrieved on February 3, 2015 from
  \textsf{www.mplayerhq.hu/design7/info.html}.

\bibitem{Man:Top}
Anonymous.
\newblock Top - display linux processes, 2015.
\newblock Manual page obtained through the command ``man top'' on {Linux}
  systems.

\bibitem{eps262294}
Marco Carbone, Mogens Nielsen, and Vladimiro Sassone.
\newblock A formal model for trust in dynamic networks.
\newblock In {\em 1st International Conference on Software Engineering and
  Formal Methods {(SEFM} 2003), 22-27 September 2003, Brisbane, Australia},
  page~54. {IEEE} Computer Society, 2003.

\bibitem{as-trust2011}
Jian Chang, Krishna~K. Venkatasubramanian, Andrew~G. West, Sampath Kannan,
  Boon~Thau Loo, Oleg Sokolsky, and Insup Lee.
\newblock {AS-TRUST:} {A} trust quantification scheme for autonomous systems in
  {BGP}.
\newblock In Jonathan~M. McCune, Boris Balacheff, Adrian Perrig, Ahmad{-}Reza
  Sadeghi, Angela Sasse, and Yolanta Beres, editors, {\em Trust and Trustworthy
  Computing - 4th International Conference, {TRUST} 2011, Pittsburgh, PA, USA,
  June 22-24, 2011. Proceedings}, volume 6740 of {\em Lecture Notes in Computer
  Science}, pages 262--276. Springer, 2011.

\bibitem{ClarkeCX09}
Stephen Clarke, Bruce Christianson, and Hannan Xiao.
\newblock Trust*: Using local guarantees to extend the reach of trust.
\newblock In Bruce Christianson, James~A. Malcolm, Vashek Matyas, and Michael
  Roe, editors, {\em Security Protocols XVII, 17th International Workshop,
  Cambridge, UK, April 1-3, 2009. Revised Selected Papers}, volume 7028 of {\em
  Lecture Notes in Computer Science}, pages 171--178. Springer, 2009.

\bibitem{DFB12b}
V.~{De~Florio} and C.~Blondia.
\newblock Safety enhancement through situation-aware user interfaces.
\newblock In {\em System Safety, incorporating the Cyber Security Conference
  2012, 7th IET International Conference on}, pages 1--6, Oct 2012.

\bibitem{De10}
Vincenzo De~Florio.
\newblock Software assumptions failure tolerance: Role, strategies, and
  visions.
\newblock In Antonio Casimiro, Rog\'erio de~Lemos, and Cristina Gacek, editors,
  {\em Architecting Dependable Systems VII}, volume 6420 of {\em Lecture Notes
  in Computer Science}, pages 249--272. Springer Berlin / Heidelberg, 2010.
\newblock 10.1007/978-3-642-17245-8\_11.

\bibitem{DF14a}
Vincenzo {De~Florio}.
\newblock Antifragility = elasticity + resilience + machine learning. models
  and algorithms for open system fidelity.
\newblock {\em Procedia Computer Science}, 32:834--841, 2014.
\newblock 1st ANTIFRAGILE workshop (ANTIFRAGILE-2015), the 5th International
  Conference on Ambient Systems, Networks and Technologies (ANT-2014).

\bibitem{DB07a}
Vincenzo De~Florio and Chris Blondia.
\newblock Reflective and refractive variables: A model for effective and
  maintainable adaptive-and-dependable software.
\newblock In {\em Proc. of the 33rd EUROMICRO Conference on Software
  Engineering and Advanced Applications (SEAA 2007)}, L{\"u}beck, Germany,
  August 2007.

\bibitem{DB07d}
Vincenzo De~Florio and Chris Blondia.
\newblock Reflective and refractive variables: A model for effective and
  maintainable adaptive-and-dependable software.
\newblock In {\em Proceedings of the 33rd Euromicro Conference on Software
  Engineering and Advanced Applications (SEEA 2007), Software Process and
  Product Improvement track (SPPI)}, L{\"u}beck, Germany, August 2007. IEEE
  Computer Society.

\bibitem{DeBl08a}
Vincenzo De~Florio and Chris Blondia.
\newblock On the requirements of new software development.
\newblock {\em International Journal of Business Intelligence and Data Mining},
  3(3), 2008.

\bibitem{DeDe98}
Vincenzo De~Florio, Geert Deconinck, Mario Truyens, Wim Rosseel, and Rudy
  Lauwereins.
\newblock A hypermedia distributed application for monitoring and
  fault-injection in embedded fault-tolerant parallel programs.
\newblock In {\em Proc. of the 6th Euromicro Workshop on Parallel and
  Distributed Processing (Euro-PDP'98)}, pages 349--355, Madrid, Spain, January
  1998. IEEE Comp. Soc. Press.

\bibitem{Festus}
Paulus~Winfridus Diaconus.
\newblock {\em Excerpta ex libris {Pompeii Festi} de significatione verborum}.
\newblock W. M. Lindsay, 1930.

\bibitem{VD13}
Jens~Van Duyse.
\newblock A toolkit for the concurrent analysis and adaptation of graphical
  user interfaces.
\newblock Master's thesis, Department of Mathematics and Computer Science,
  University of Antwerp, Middelheimlaan 1, 2020 Antwerpen, Belgium, 2013.
\newblock Promotor: {Vincenzo De~Florio}.

\bibitem{5340804}
Tyrone Grandison and Morris Sloman.
\newblock A survey of trust in internet applications.
\newblock {\em Commun. Surveys Tuts.}, 3(4):2--16, October 2000.

\bibitem{GRT}
Michael Grottke, Matias~Jr. Rivalino, and Kishor~S. Trivedi.
\newblock The fundamentals of software aging.
\newblock In {\em Proceedings of the 1st International Workshop on Software
  Aging\ and Rejuvenation, 19 International Symposium on Software Reliability
  Engineering}, 2008.

\bibitem{Assets:GrTr07}
Michael Grottke and Kishor~S. Trivedi.
\newblock Fighting bugs: Remove, retry, replicate, and rejuvenate.
\newblock {\em {IEEE} Computer}, 40(2):107--109, 2007.

\bibitem{primiero_NDC}
Giuseppe~Primiero Jaap~Boender and Franco Raimondi.
\newblock Minimizing transitive trust threats in software management systems.
\newblock Technical report, Foundations of Computing Group, Middlesex
  University, 2015.

\bibitem{MAPE}
Bart Jacob, Richard Lanyon-Hogg, Devaprasad~K Nadgir, and Amr~F Yassin.
\newblock {\em A Practical Guide to the IBM Autonomic Computing Toolkit}.
\newblock IBM Redbooks, 2004.

\bibitem{Therac93}
Nancy Leveson and Clark~S. Turner.
\newblock An investigation of the {Therac-25} accidents.
\newblock {\em IEEE Computer}, 26(7):18--41, 1993.

\bibitem{primiero_secureND}
Giuseppe Primiero and Franco Raimondi.
\newblock A typed natural deduction calculus to reason about secure trust.
\newblock In Ali Miri, Urs Hengartner, Nen{-}Fu Huang, Audun J{\o}sang, and
  Joaqu{\'{\i}}n Garc{\'{\i}}a{-}Alfaro, editors, {\em 2014 Twelfth Annual
  International Conference on Privacy, Security and Trust, Toronto, ON, Canada,
  July 23-24, 2014}, pages 379--382. {IEEE}, 2014.

\bibitem{primiero_taddeo}
Giuseppe Primiero and Mariarosaria Taddeo.
\newblock A modal type theory for formalizing trusted communications.
\newblock {\em J. Applied Logic}, 10(1):92--114, 2012.

\bibitem{PatriotNYT}
Eric Schmitt.
\newblock After the war; army is blaming patriot's computer for failure to stop
  the dhahran scud.
\newblock {\em New York Times}, May 1991.

\bibitem{journals/tdsc/YanP11}
Zheng Yan and Christian Prehofer.
\newblock Autonomic trust management for a component-based software system.
\newblock {\em IEEE Transactions on Dependable and Secure Computing},
  8(6):810--823, 2011.

\end{thebibliography}
\bibliographystyle{plain}




\end{document}